# Impact of random spatial truncation and reciprocal-space binning on the detection of hyperuniformity in disordered systems

Yuan Liu (刘源) ; Xurui Li (李旭瑞) ; Jianxiang Tian (田建祥) ; Xunwang Yan (闫循旺) ; Ge Zhang (张格)

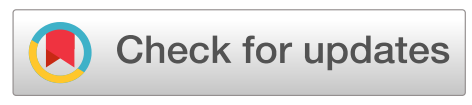

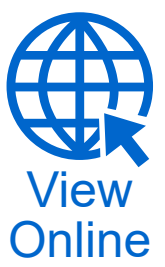

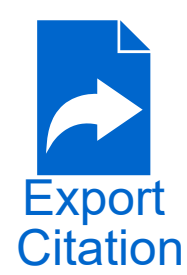

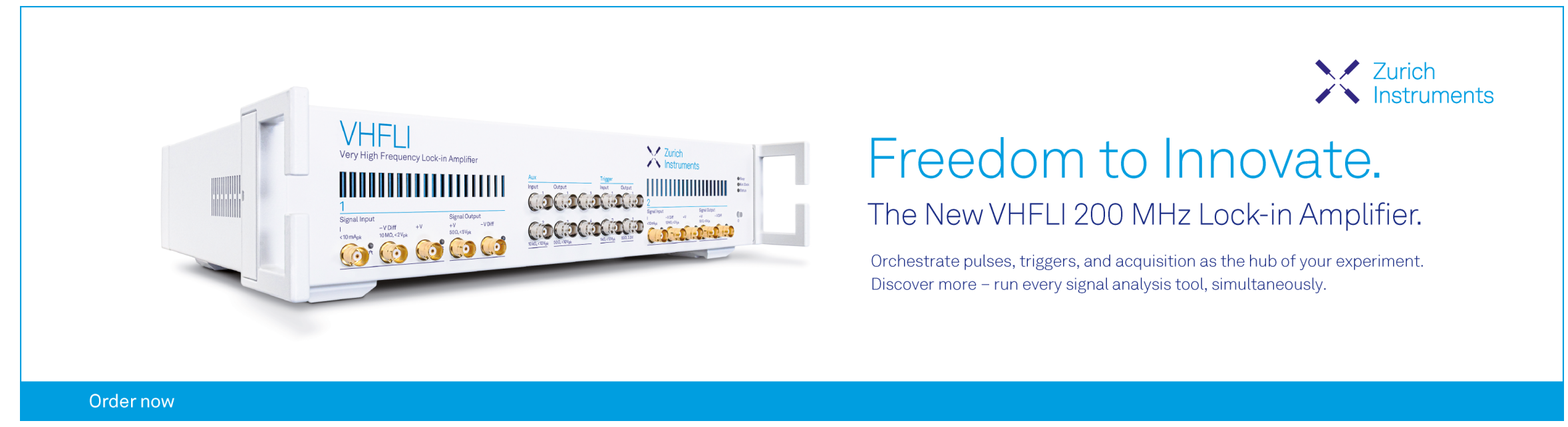

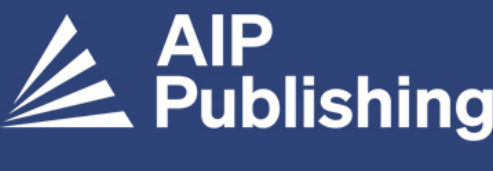

# Impact of random spatial truncation and reciprocal-space binning on the detection of hyperuniformity in disordered systems

Yuan Liu (刘源),[1] Xurui Li (李旭瑞),[1] Jianxiang Tian (田建祥),[1,2,a)] Xunwang Yan (闫循旺),[1,a)] and Ge Zhang (张格)[3,a)]

## AFFILIATIONS

[1] Department of Physics, Qufu Normal University, Qufu 273165, People's Republic of China
[2] Department of Physics, Dalian University of Technology, Dalian 116024, People's Republic of China
[3] Department of Physics, City University of Hong Kong, Kowloon, Hong Kong, China

[a)] **Authors to whom correspondence should be addressed:** jxtian@qfnu.edu.cn; xwyan@qfnu.edu.cn; and gzhang37@cityu.edu.hk

## ABSTRACT

We study how finite-window sampling (random spatial truncation) and reciprocal-space radial binning influence the detection of hyperuniformity in disordered systems. Starting from thirteen representative two-dimensional simulation systems and two experimental biological systems, we apply random spatial truncation and then rescale the cropped systems to a fixed number density. We then compute the structure factor $S(k)$ and the local number variance $\sigma_N^2(R)$ to determine whether cropped configurations preserve the salient structural properties of the original ones. We find that moderate random spatial truncation does not change qualitatively the hyperuniformity classification of the systems, despite a reduction in measured hyperuniformity exponent $\alpha$ for $\alpha > 1$. Since spatial truncations exacerbate fluctuations in measured $S(k)$ at small $k$, we show that modest reciprocal-space radial binning (controlled by a binning parameter $m$) effectively smooths out such fluctuations without changing the hyperuniformity class. Practical guidelines for choosing $m$ and cross-checking $S(k)$ fits with $\sigma_N^2(R)$ scaling are provided. Our research provides a concrete, low-cost, and effective methodology for robust detection and classification of hyperuniformity in finite and truncated datasets, which are prevalent in experimental systems.

## I. INTRODUCTION

Hyperuniform systems constitute a class of many-body systems with distinctive structural correlations, characterized by an anomalous suppression of density fluctuations in the long-wavelength (large-scale) limit. Hyperuniform systems include perfect crystals, perfect quasicrystals, and certain special disordered systems, while non-hyperuniform systems include typical gases, liquids, and glasses.[1] The concept of hyperuniformity, therefore, provides a unified framework for the classification and structural characterization of crystals, quasicrystals, and those special disordered systems.[2–4]

Research on hyperuniformity has recently emerged as an interdisciplinary field and is being actively pursued in diverse directions, including photonic and phononic bandgap materials,[5–9] antenna and laser design,[10] stealthy hyperuniform materials,[11,12] transport phenomena and critical currents in superconductors,[13] diffusion in two-phase media,[14–17] properties in condensed matter,[18] point sets in pure mathematics,[19–22] network systems,[23] epithelial cells,[24] classical and quantum spin liquids,[25] self-assembled robotic spinners,[26] chromatophores on squid skin,[27] anisotropic random fields,[28] jammed packings,[29] frustrated Vicsek–Kuramoto systems,[30] and the complex dynamic behaviors induced by cross-layer interactions in multilayer systems.[31] These studies have found extensive applications in the design and optimization of novel disordered hyperuniform materials.[32–38]

As investigations into the relationship between microscopic structure and function in biological systems deepen, the significance of hyperuniformity has also become increasingly evident.

In avian retinae, the photoreceptor cones exhibit a hyperuniform spatial organization. By suppressing long-wavelength density fluctuations, this arrangement enhances photon-capture efficiency and optimizes signal encoding, thereby supporting robust visual perception under complex environmental conditions.[39] In immune systems, the self-organization of immune cells into hyperuniform patterns likewise promotes efficient spatial allocation of cell types and facilitates coordinated interactions, improving the precision and efficacy of immune responses.[40] At the ecosystem scale, hyperuniform vegetation patterns observed in arid and semi-arid regions are interpreted as adaptive responses to limited water availability and poor soils. Such patterns reduce local competition, optimize resource exploitation, and help stabilize material cycling and energy flows, with important consequences for ecosystem resilience and function.[41] In addition, a global study on drylands found that disordered hyperuniformity shapes vegetation patterns in roughly one-tenth of drylands. These patterns appear visually disordered on small scales while maintaining consistent uniformity across larger scales. This self-organization stems from plant–plant or plant–sediment interactions. It enhances water retention, expands the system's aridity tolerance, and connects to diverse dryland systems, although it may impede recovery following disturbances.[42] Moreover, leaf vein networks, which are hierarchical self-organized architectures, exhibit disordered hyperuniformity. Geometric constraints from primary veins and branching patterns prevent the veins from evolving into a perfectly ordered lattice; thus, the resulting disordered hyperuniform topology achieves near-optimal diffusion and transport.[43]

In both simulations and experiments, the most prevalent method to detect hyperuniformity is to compute the structure factor, $S(\boldsymbol{k})$, whose small-$\boldsymbol{k}$ behavior encodes the system's large-scale spatial correlations. For an infinite, statistically homogeneous configuration with number density $\rho$, $S(\boldsymbol{k})$ is related to the pair correlation function $g_2(\boldsymbol{r})$ by

$$S(\boldsymbol{k}) = 1 + \rho \tilde{h}(\boldsymbol{k}), \tag{1}$$

where $\tilde{h}(\boldsymbol{k})$ denotes the Fourier transform of the total correlation function $h(\boldsymbol{r}) = g_2(\boldsymbol{r}) - 1$. For isotropic systems, $S(\boldsymbol{k})$ only depends on the scalar wavenumber $k = |\boldsymbol{k}|$. A configuration is hyperuniform if the structure factor vanishes in the infinite-wavelength limit,

$$\lim_{k\to 0} S(k) = 0, \tag{2}$$

which implies the suppression of density fluctuations at very large length scales.[3] For finite samples (e.g., a simulation box of linear size $L$ with $N$ points under periodic boundary conditions, or a finite experimental field of view), one computes the configuration-specific structure factor from the particle positions $\{\boldsymbol{r}_1, \ \boldsymbol{r}_2, \ldots, \boldsymbol{r}_N\}$,[4]

$$S(\boldsymbol{k}) = \frac{1}{N}\left|\sum_{j=1}^{N} e^{i\boldsymbol{k}\cdot\boldsymbol{r}_j}\right|^2 \quad (\boldsymbol{k} \neq 0), \tag{3}$$

and then performs ensemble averaging and angular averaging as appropriate. Since the forward-scattering term at $\boldsymbol{k} = 0$ is excluded in this equation, and since the allowed discrete wavenumbers are $\boldsymbol{k} = \boldsymbol{n}\cdot(2\pi/L)$ (with integer vector $\boldsymbol{n}$), one cannot study arbitrarily small $k$ with finite systems. Therefore, reliable assessment of $\lim_{k\to 0} S(k)$ requires extrapolation from the smallest accessible $k$ values, together with careful control of finite-size effects and sampling noise.

An equivalent, complementary real-space method to detect hyperuniformity is to compute the local number variance $\sigma_N^2(R) \equiv \langle N^2(R)\rangle - \langle N(R)\rangle^2$, which measures fluctuations of the number of particles, $N$, inside an observation window of characteristic linear size $R$. The mean of $N(R)$ is $\langle N(R)\rangle = \rho v_1(R)$, where $v_1(R)$ is the window volume, for statistically homogeneous systems. The variance of $N(R)$ was analytically found to be

$$\sigma_N^2(R) = \rho v_1(R)\left[1 + \rho\int_{\mathbb{R}^d} h(\boldsymbol{r})\alpha_2(\boldsymbol{r};R)d\boldsymbol{r}\right], \tag{4}$$

where $\alpha_2(\boldsymbol{r};R)$ denotes the normalized intersection volume of two identical windows whose centers are separated by $|r|$, if windows are spherical.[3] One can also derive a Fourier-space representation for Eq. (4),

$$\sigma_N^2(R) = \langle N(R)\rangle\frac{1}{(2\pi)^d}\int_{\mathbb{R}^d} S(\boldsymbol{k})\tilde{\alpha}_2(\boldsymbol{k};R)d\boldsymbol{k}, \tag{5}$$

where $\tilde{\alpha}_2(\boldsymbol{k};R)$ is the Fourier transform of $\alpha_2(\boldsymbol{r};R)$.[3] This identity shows that the suppression of $S(\boldsymbol{k})$ at small $\boldsymbol{k}$ necessarily reduces the contribution of long-wavelength modes to $\sigma_N^2(\boldsymbol{R})$ and conversely that a $\sigma_N^2(\boldsymbol{R})$ that grows slower than $R^d$ for large $R$ implies vanishing $S(\boldsymbol{k})$ at $k\to 0$. Hence, $S(\boldsymbol{k})$ and $\sigma_N^2(\boldsymbol{R})$ provide two mathematically equivalent and, therefore, complementary, mutually reinforcing methods for detecting hyperuniformity.[44–47]

When $S(k)$ vanishes as a power law,

$$S(k) \sim k^\alpha \quad (k\to 0), \tag{6}$$

hyperuniform disordered systems can be further divided into three classes. Here, $\alpha$ is referred to as the hyperuniformity exponent.[1] In each class, the asymptotic scaling of the local number variance $\sigma_N^2(R)$ can also be analytically derived,[3,4,48] as detailed in Table I. Besides using the large-$R$ leading term of $\sigma_N^2(R)$ to classify hyperuniformity, its finite-$R$ behavior has been used to quantify the degree of order/disorder,[49] while its large-$R$ subleading corrections were also found to be dependent on the hyperuniformity class.[50]

From a practical standpoint, reliable identification of disordered hyperuniformity in small systems often requires a combined analysis of $S(k)$ and $\sigma_N^2(R)$, together with careful control of multiple potential biases: Finite simulation boxes or observation windows impose a discrete $k$-grid, forcing one to perform extrapolations with quantified uncertainties in order to access the $k\to 0$ regime.

**TABLE I.** Classification of hyperuniform systems by the small-$k$ scaling $S(k) \sim k^\alpha$ and the corresponding asymptotic growth of the local number variance $\sigma_N^2(R)$ in $d$ dimensions.

| Hyperuniformity class | Value of $\alpha$ | Asymptotic scaling of $\sigma_N^2(R)$ vs $R$ |
|---|---|---|
| Class I | $\alpha > 1$ | $\sigma_N^2(R) \sim R^{d-1}$ |
| Class II | $\alpha = 1$ | $\sigma_N^2(R) \sim R^{d-1}\ln\ (R)$ |
| Class III | $0 < \alpha < 1$ | $\sigma_N^2(R) \sim R^{d-\alpha}$ |

Moreover, spatial truncation and edge effects may introduce systematic error and fluctuations in $S(k)$, which can be clearly observed in experiments.[39–41,43] Related finite-window challenges also arise in spectral assessments of disordered hyperuniformity in two-dimensional materials and networks based on atomic-scale micrographs.[51,52] Conversely, the study of $\sigma_N^2(R)$ is also limited by the system size, preventing the access of the infinite-$R$ limit.

In this paper, we systematically explore how spatial truncation and related sampling limitations affect the structure factor $S(k)$ and the local number variance $\sigma_N^2(R)$ in controlled simulation systems, develop a practical framework for detecting and classifying hyperuniformity in datasets with limited system sizes, and demonstrate the effectiveness of our framework on real-life biological systems with a limited field of view. We show how the recommended procedures mitigate truncation-induced artifacts and improve the reliability of hyperuniformity assessments. Although finite samples cannot completely reproduce infinite-system phenomena, we show that the exponent $\alpha$ is biased in a systematic (reproducible) way, and the hyperuniformity class is preserved.

The remainder of this paper is organized as follows: In Sec. II, we introduce the controlled simulation systems and experimental biological systems and present the protocols for random spatial truncation, rescaling, and reciprocal-space radial binning used in our analyses. Our results are presented in Sec. III. In particular, Sec. III A presents an overview of spatial truncation effects across thirteen representative two-dimensional controlled simulation systems. Section III B examines the effects of the crop size on spectral and real-space protocols for detecting hyperuniformity. Section III C studies the influence of reciprocal-space radial binning, demonstrates how moderate binning mitigates small-$k$ fluctuations, and provides practical guidance for selecting the degree of binning. Section III D applies the developed protocol to experimental biological systems (avian photoreceptor mosaics and looped leaf vein networks) and shows that modest radial binning smooths out small-$k$ artifacts without changing the inferred hyperuniform classification. In Sec. IV, we present conclusions and outline directions for future research.

## II. METHODS

To explore how spatial truncation affects the detection of hyperuniformity in disordered systems by altering the structure factor $S(k)$ and the local number variance $\sigma_N^2(R)$,[1] we analyzed thirteen controlled simulation systems together with two classes of experimental biological systems. These systems were chosen to span a broad range of spatial correlations and practical data conditions, thereby enabling us to test the effects of spatial truncation under realistic constraints. The number of configurations is $N_c = 20$ so that conclusions apply directly to experimental datasets of comparable sizes. For consistency, all systems are two-dimensional, and every configuration is rescaled to a target number density of $\rho = N/L^2 = 1.5$ unless otherwise specified, where $N$ is the number of particles and $L$ is the side length of the simulation box. The structure factors $S(k)$, pair correlation functions $g_2(r)$, and local number variances $\sigma_N^2(R)$ are computed in periodic boxes after rescaling. Representative configurations of all controlled simulation systems are shown in Fig. 1, while those for experimental biological systems are given in Refs. 39 and 43.

### A. Controlled simulation systems

We employed thirteen typical two-dimensional point-pattern models as controlled testbeds.

(1) **Stealthy hyperuniform systems:** Stealthy hyperuniform systems are disordered ground states of stealthy pair interactions and exhibit $S(\boldsymbol{k}) = 0$ for every $|\boldsymbol{k}|$ below a prescribed cutoff. They correspond to the $\alpha \to \infty$ limit in $S(k) \sim k^{\alpha}$ and, therefore, belong to class I hyperuniformity. These patterns interpolate between uncorrelated systems (ideal gases) and crystals as the constraint parameter, $\chi$, increases.[53–55] We generated these ground states by energy minimization and studied two representative stealthy hyperuniform systems with $\chi = 0.20$ (low correlation) and $\chi = 0.49$ (high correlation) to represent weakly and strongly correlated stealthy states.

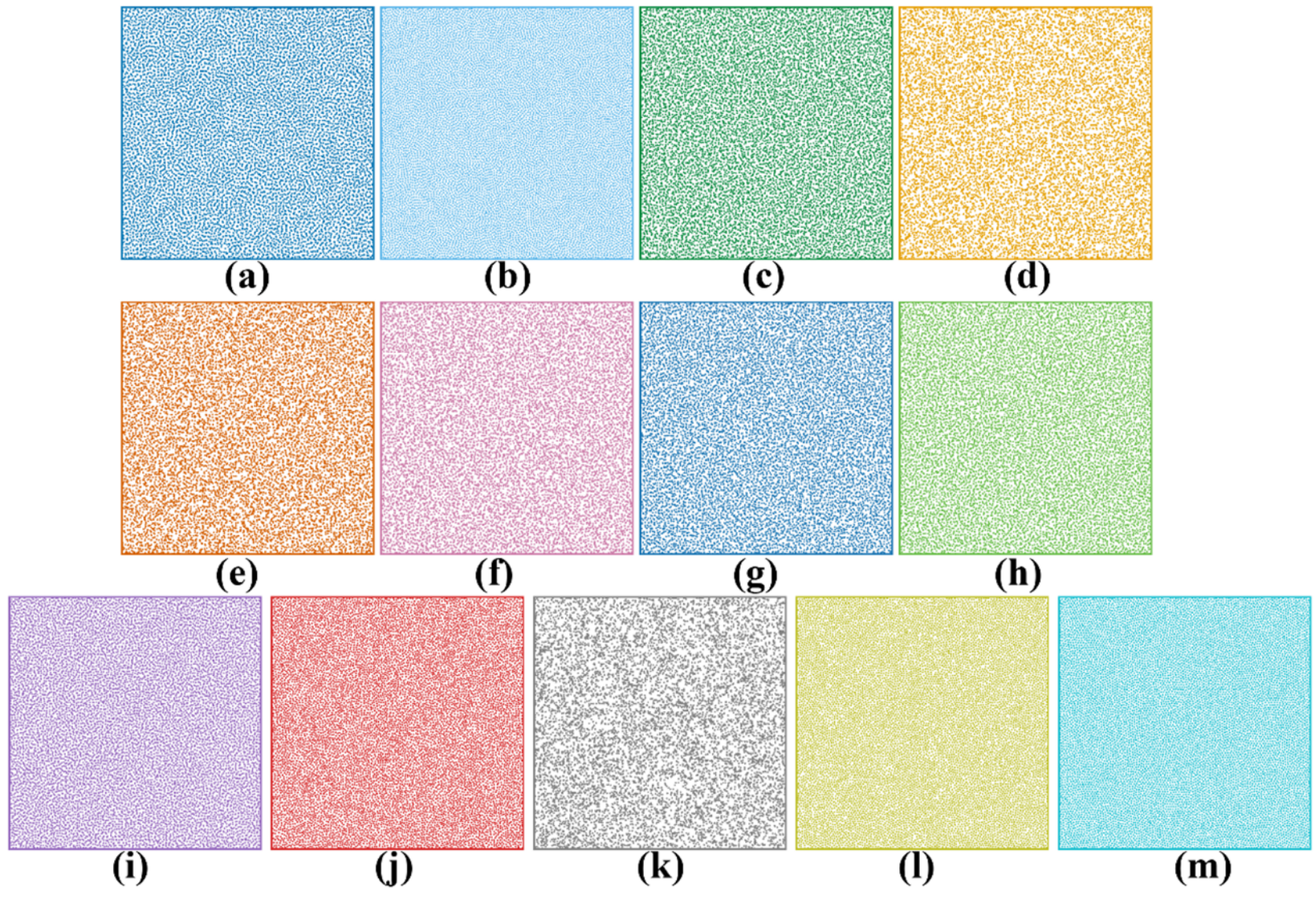

**FIG. 1.** Representative configurations of the controlled simulation systems used in this study. Each panel shows a 2D configuration of $N = 10^4$ points. (a) and (b) Stealthy hyperuniform systems with $\chi = 0.20$ and $\chi = 0.49$, respectively; (c) hyperuniform Gaussian pair statistics system; (d)–(i) hyperuniform targeted $S(k)$ systems with targeted $\alpha = 0.5$, 0.7, 1.0, 1.3, 1.5, and 3.0, respectively; (j) random sequential addition (RSA) system; (k) Poisson point distribution system; (l) Lennard-Jones fluid system; and (m) one-component Yukawa fluid system.

(2) **Hyperuniform Gaussian pair statistics system:** The equilibrium state of the 2D Coulombic interaction possesses a total correlation function $h(r)$ that is a Gaussian with an amplitude of $-1$, and so, its Fourier transform is likewise a Gaussian. The structure factor of this system vanishes at the origin with the power-law $S(k) \sim k^2$, corresponding to class I hyperuniformity. Because this system provides a concrete, realizable example of disordered hyperuniformity with a hyperuniformity exponent $\alpha = 2$, it serves as a convenient canonical reference for benchmarking sampling limitations.[56]

(3) **Hyperuniform targeted $S(k)$ systems:** Systems satisfying the relationship $S(k) \sim k^\alpha$ were generated using the algorithm detailed in Ref. 57. In particular, this algorithm constructs 100 configurations simultaneously by minimizing the difference between their average structure factor and a target, $S_0(k) = (k/k_{\max})^{\alpha}$, for all $|\boldsymbol{k}| < k_{\max}$. Here, we chose $k_{\max} = 6.1$ so that the target is realizable at $\rho = 1.5$ for every $\alpha$ we explore. We selected six values of $\alpha$: 0.5, 0.7, 1.0, 1.3, 1.5, and 3.0, which fully cover the three hyperuniformity classes, in order to systematically study sampling limitations.

(4) **Random Sequential Addition (RSA) system:** RSA packings are generated by irreversible, non-overlapping adsorptions of disks on a plane, which approach a saturation density with characteristic short-range exclusion and correlations. Near-saturation RSA exhibits suppressed local clustering relative to a completely random process but is not hyperuniform. RSA systems serve as a nonequilibrium reference with strong local order.[58,59] For this study, we employed near-saturation RSA configurations that met two specific criteria: one being 10 000 consecutive failed attempts to insert new particles, while the other being an exact particle count of 10 000.

(5) **Poisson point distribution system:** The homogeneous Poisson point distribution system (ideal gas) serves as the null model, i.e., a complete lack of spatial order. In the infinite-size limit, it has $S(k) \equiv 1$ and $\sigma_N^2(R) \sim R^d$, i.e., no large-scale suppression at all. It is included as a non-hyperuniform reference system.[60]

(6) **Lennard–Jones (LJ) fluid system:** Depending on the state point, an LJ fluid can show mild small-$k$ suppression and interparticle correlations but is not hyperuniform. LJ fluids, therefore, provide an equilibrium liquid benchmark with a realistic short-range structure and thermal noise.[61,62] We employed reduced density $\rho^* = \rho_0\sigma^3 = 0.8$ and reduced inverse temperature $1/T^* = \epsilon/(k_B T) = 1.35$, where $\sigma$ represents the distance between two particles when the LJ potential is exactly zero, $\epsilon$ is the depth of the LJ potential well, $k_B$ stands for the Boltzmann constant, $\rho_0$ represents the density, and $T$ denotes the temperature.[61] We generated configurations of the LJ fluid system via molecular dynamics (MD) simulations using the Verlet algorithm.[63] The initial configuration was set to a square lattice, and configurations were periodically sampled every 10 000 time steps. The reduced time step was set to $\Delta t^* = 0.001$ (where the reduced time is defined as $t^* = t\sqrt{\epsilon/m}/\sigma$), and the maximum reduced time was $t^*_{\max} = 500$. All configurations were based on the converged results after $5 \times 10^5$ time steps, ensuring that the system reached full equilibrium.

(7) **One-component Yukawa fluid system:** Yukawa (screened-Coulomb) systems model charged colloids and dusty plasmas. By tuning screening and coupling parameters, they exhibit a wide range of compressibility and correlation strengths. Under strong coupling and weak screening, these systems can display substantial small-$k$ suppression. Nevertheless, they are not hyperuniform.[64,65] We also employed the same reduced density and reduced temperature as the LJ fluid system. Configurations of the one-component Yukawa fluid were generated using the same MD simulation protocol as the LJ fluid. In particular, we employed the same initial configuration, sampling interval, reduced time step, maximum reduced time, and equilibration time, all to ensure consistency across the two fluid models.

## B. Experimental biological systems

To demonstrate practical implications, we also analyzed two classes of experimental biological systems.

(1) **Avian photoreceptor patterns (bird retina cones):** Five subtypes of photoreceptors (cones) exist in bird retinas, and they constitute a striking example of biological multihyperuniformity, that is, both the combined cone population and each individual cone subtype exhibit vanishing $S(k)$ as $k \to 0$. Such patterns are hypothesized as a result of optimizing light sampling under the constraint of packing polydisperse cells. We use published coordinate datasets for chicken retinae as a biologically motivated empirical test case.[39]

(2) **Looped leaf vein networks:** Recent analyses of looped secondary vein networks demonstrate a universal hyperuniform organization of the cellular loop centers (a biological network tessellation), with measured small-$k$ scaling consistent with disordered hyperuniformity. Leaf vein thus provides another structurally distinct real-world example in which truncated observations and finite sampling are genuine practical issues.[43]

## C. Definitions of cropping ratio parameter and radial binning parameter

To investigate how spatial truncation and finite sampling affect the structure factor $S(k)$ and the local number variance $\sigma_N^2(R)$, we introduce two parameters, $c$ and $m$, detailed below. All spatial truncation is performed by random square cropping. Starting from a configuration in a square box of side $L$, we choose a square crop of linear size $cL$ with $0 < c \leq 1$. The crop center is chosen randomly with uniform distribution. All points with positions inside the crop are kept, yielding a cropped point set with a particle count of $N_{crop}$. To preserve the target number density $\rho$ for subsequent analysis, the cropped configuration is rescaled into a periodic square system of side length $L_{crop} = \sqrt{N_{crop}/\rho}$. Thus, $c = 1.0$ corresponds to no cropping with $L_{crop} = L$, but for $c < 1$, the cropped side length $L_{crop}$ can fluctuate around its expectation of $cL$.

Typically, in a periodic box of side, when calculating the structure factor $S(k)$, the allowed wave numbers are $k = \boldsymbol{n} \cdot (2\pi/L)$ (with integer vector $\boldsymbol{n}$) so that a natural radial bin width is $2\pi/L$.[1] To control radial smoothing of the measured structure factor, we introduce a reciprocal-space radial binning parameter $m$ and partition the radial $k$-axis into uniform bins of width $k_{Bin} = m \cdot (2\pi/L)$. Thus, $m$ controls the number of discrete reciprocal vectors pooled into each bin. Notably, wavevector quantization under periodic boundary conditions and finite-size limitations make $m \geq 1.0$ necessary.[66,67] In practice, $m$ is treated as a numerical regularization parameter whose influence is examined as part of the analysis.

### D. Determining exponents $\alpha$ and $\beta$

As explained in the Introduction, hyperuniformity can be detected and classified by studying either the small-$k$ scaling of $S(k)$ [$\alpha$ in $S(k) \propto k^{\alpha}$] or the large-$R$ scaling of $\sigma_N^2(R)$ [$\beta$ in $\sigma_N^2(R) \propto R^{\beta}$]. For smaller systems, one often needs to study both exponents so that they complement each other to provide better assessments. For the estimation of $\alpha$, we performed linear least-square regression of $\ln[S(k)]$ vs $\ln k$ and extracted the slope. The fit is restricted to a small $k$ range of $ka \in (0, 2.5]$, where $a$ is the ensemble-averaged nearest-neighbor distance.

For the estimation of $\beta$, we applied the same procedure, performing a linear fit of $\ln[\sigma_N^2(R)]$ vs $\ln R$ and extracting the slope. The exponent $\beta$ only applies to the large-$R$ limit, but previous research discovered that $\sigma_N^2(R)$ suffers from significant finite-size effects for $R > L/4$.[68] We, therefore, let the fitting window be $R \in [0.175L, 0.2L]$, with a sampling step size of $\Delta R = 0.005L$.

## III. RESULTS

### A. Spatial truncation effects in controlled simulation systems

To evaluate the generic consequences of spatial truncation, we compare spectral and real-space statistics across thirteen representative two-dimensional controlled simulation systems. For each system, we generated configurations with $N = 10\,000$ particles, applied random spatial truncation (i.e., randomly extracting a smaller subwindow from the original configuration, as illustrated in Fig. 2) with ratio $c = 0.1$, rescaled the cropped configurations to the same number density (see Sec. II C), and then computed the structure factor $S(k)$, the local number variance $\sigma_N^2(R)$, and the pair correlation function $g_2(r)$. Figure 2 presents these quantities before and after truncation for all thirteen systems, while Table II summarizes the realized mean cropped particle number $\overline{N}_{crop}$, fitted hyperuniformity exponent $\alpha$, the difference in the fitted hyperuniformity exponent before and after cropping $\Delta\alpha$, and fitted local number variance exponent $\beta$ for the uncropped and cropped systems.

It is important to note that the theoretical values of exponents $\alpha$ and $\beta$ for each disordered hyperuniform system can be achieved only from ensemble averaging, while those obtained from our limited sample of 20 configurations can fluctuate. In addition, discrete-$k$ resolution, the choice of radial binning, and errors associated with numerical simulation and fitting processes also induce moderate deviations. As a prominent example of the aforementioned inaccuracies, class II hyperuniform systems are defined as $\alpha$ strictly equal to 1, but the fitted $\alpha$ for such systems can vary in the range of 0.95–1.05 in this paper.

It can be observed from Table II that for hyperuniform systems (stealthy systems, hyperuniform Gaussian pair statistics systems, and hyperuniform targeted $S(k)$ systems), the characteristic small-$k$ suppression of $S(k)$ is retained following cropping, but cropping causes two reproducible effects. First, the hyperuniformity exponent $\alpha$ obtained from cropped configurations is systematically reduced compared with the uncropped case, which indicates a modest weakening of long-wavelength suppression. Systems with a higher $\alpha$ prior to cropping exhibit a greater reduction in α after cropping. Although this reduction is significant for $\alpha = 3$, it diminishes and becomes negligible for $\alpha \leq 1$, and, therefore, does not alter the asymptotic hyperuniform class assignment of the system. Second, cropping at $c = 0.1$ tends to amplify spectral fluctuations in the very smallest $k$-bins, inducing nonmonotonic features (wiggles) that may bias small-$k$ fits. Apart from these, the short-range structure is robust to spatial truncation, and the overall shapes of $g_2(r)$ and $S(k)$ are preserved. This feature is present across both hyperuniform and non-hyperuniform systems.

For non-hyperuniform systems, Poisson systems remain unchanged [$S(k) \approx 1$] before and after cropping, up to sampling noise. The random sequential addition (RSA) system, the Lennard-Jones (LJ) fluids, and the Yukawa fluids at the chosen liquid state points all exhibit nonzero small-$k$ intercepts in the uncropped state and retain these intercepts after cropping. Therefore, spatial truncation does not drive these systems toward hyperuniformity.

The results of local number variance $\sigma_N^2(R)$ corroborate the structure factor findings. Cropped systems preserve the characteristic scaling of $\sigma_N^2(R)$, as class I and II hyperuniform systems maintain $\beta \approx 1.0$, while non-hyperuniform systems retain their original $\beta$ behavior within sampling uncertainty. Taken together, the spectral and real-space results emphasize that reliable hyperuniformity classification requires joint consideration of both $S(k)$ and $\sigma_N^2(R)$.

Taken together, these results support three conclusions. First, random spatial truncation at the moderate level here does not convert non-hyperuniform systems into hyperuniform ones, since ensembles that exhibit a finite intercept in $S(k)$ before truncation continue to do so after truncation. Second, for systems that are disordered hyperuniform (i.e., that exhibit strong small-$k$ suppression prior to truncation), spatial truncation weakens the degree of suppression (manifesting as a reduction in fitted $\alpha$) but does not change the asymptotic hyperuniformity class as determined by the scaling of $S(k)$ and $\sigma_N^2(R)$. Third, spatial truncation largely preserves short-range structural information [peak positions, peak counts, and other features in $g_2(r)$ and $S(k)$]. Nevertheless, spatial truncation can induce configuration-dependent fluctuations in the first few small-$k$ bins of $S(k)$. These effects are attributable to the combined effects of loss of true long-wavelength modes and discrete-grid and rebinning artifacts that arise when $N_{crop}$ and $L_{crop}$ differ from the uncropped values. Finite-sample noise and configuration dependence further modulate observed small-$k$ behavior, which motivates the subsequent focused analyses of $c$- and $m$-dependences in Secs. III B and III C.

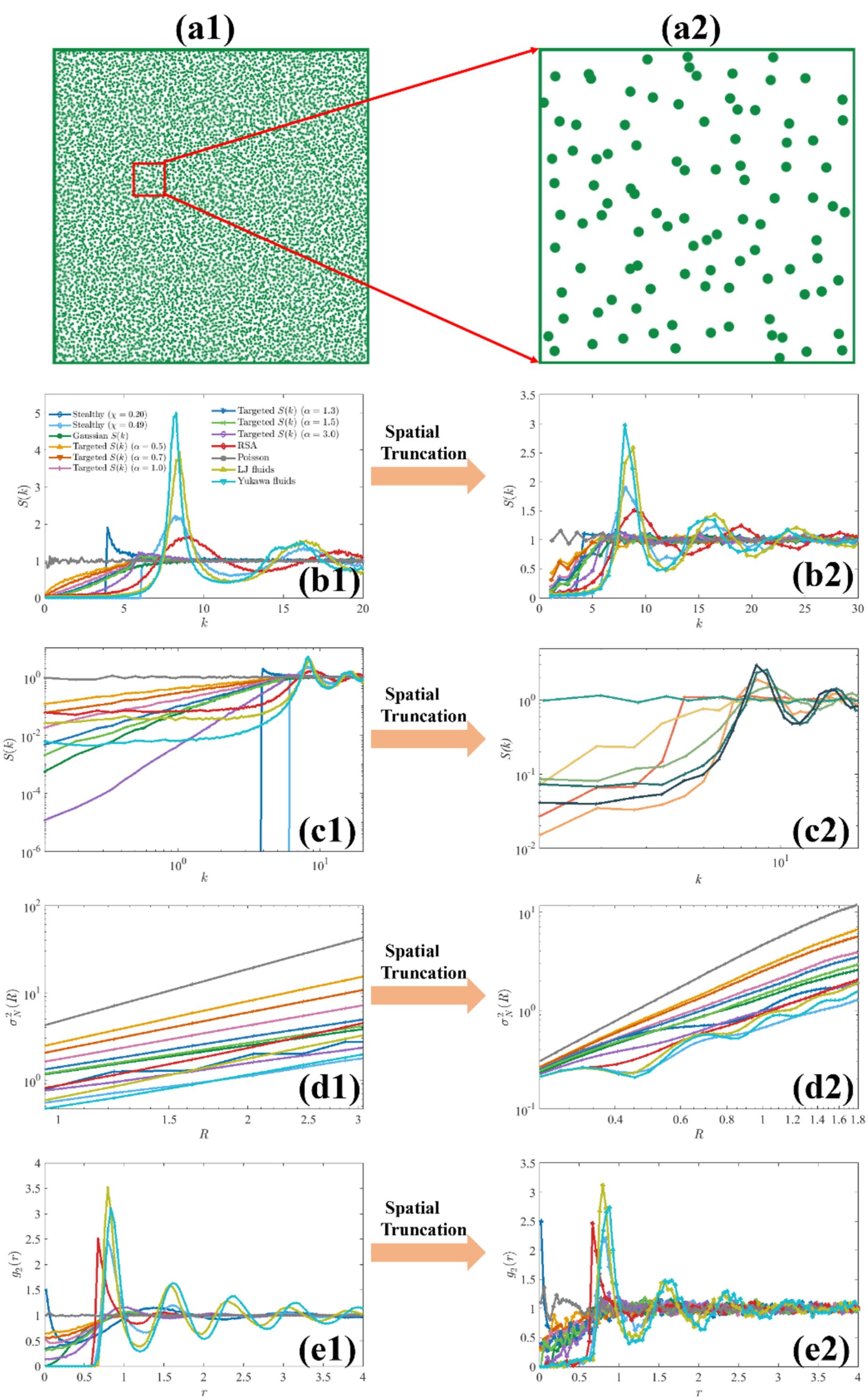

**FIG. 2.** Comparison of thirteen representative two-dimensional simulation systems before and after truncation. Left column [(a1)–(e1)] shows results for uncropped systems, while right column [(a2)–(e2)] shows results for systems after random spatial truncation with $c = 0.1$ and rescaling to the original number density. Panels [(a1) and (a2)] illustrate the process of random spatial truncation: (a1) shows the full-field configuration with $N = 10\,000$ and the red inset indicates the cropping window, and (a2) shows the corresponding cropped configuration with $N = 100$. Panels [(b1) and (b2)] show the structure factor $S(k)$ on a linear scale. Panels [(c1) and (c2)] show the same $S(k)$ data on a log–log scale. Panels [(d1) and (d2)] present the local number variance $\sigma_N^2(R)$ on a log–log scale. Panels [(e1) and (e2)] display the pair correlation function $g_2(r)$.

TABLE II. Summary of spatial truncation effects. For each model, the crop ratio $c$, the realized mean cropped particle number $\bar{N}_{crop}$, the fitted hyperuniformity exponent $\alpha$ obtained from a small-$k$ fit to $S(k)$, the difference in the fitted hyperuniformity exponent $\alpha$ before and after cropping ($\Delta\alpha$), and the fitted real-space exponent $\beta$ obtained from the asymptotic scaling of $\sigma_N^2 \sim R^\beta$ are listed. All values are reported for uncropped ($c = 1.0$) and cropped ($c = 0.1$) systems.

| System | $c$ | $\bar{N}_{crop}$ | $\alpha$ | $\Delta\alpha$ | $\beta$ |
|---|---|---|---|---|---|
| Stealthy ($\chi = 0.20$) system | 1.0<br>0.1 | 10 000<br>100.2 | $\infty$<br>1.92 | $\infty$ | 1.01<br>1.00 |
| Stealthy ($\chi = 0.49$) system | 1.0<br>0.1 | 10 000<br>99.6 | $\infty$<br>1.96 | $\infty$ | 1.02<br>1.02 |
| Hyperuniform Gaussian pair statistics system | 1.0<br>0.1 | 10 000<br>100.5 | 2.04<br>1.59 | 0.45 | 0.99<br>1.00 |
| Hyperuniform targeted $S(k)$ systems (targeted $\alpha = 0.50$) | 1.0<br>0.1 | 10 000<br>98.1 | 0.53<br>0.49 | 0.04 | 1.46<br>1.45 |
| Hyperuniform targeted $S(k)$ systems (targeted $\alpha = 0.70$) | 1.0<br>0.1 | 10 000<br>101.2 | 0.73<br>0.68 | 0.05 | 1.25<br>1.24 |
| Hyperuniform targeted $S(k)$ systems (targeted $\alpha = 1.00$) | 1.0<br>0.1 | 10 000<br>98.9 | 1.02<br>0.96 | 0.06 | 1.07<br>1.05 |
| Hyperuniform targeted $S(k)$ systems (targeted $\alpha = 1.30$) | 1.0<br>0.1 | 10 000<br>100.7 | 1.31<br>1.15 | 0.16 | 1.03<br>1.04 |
| Hyperuniform targeted $S(k)$ systems (targeted $\alpha = 1.50$) | 1.0<br>0.1 | 10 000<br>100.3 | 1.52<br>1.27 | 0.25 | 1.00<br>1.02 |
| Hyperuniform targeted $S(k)$ systems (targeted $\alpha = 3.00$) | 1.0<br>0.1 | 10 000<br>99.7 | 2.97<br>1.64 | 1.33 | 0.99<br>0.98 |
| Random sequential addition (RSA) system | 1.0<br>0.1 | 10 000<br>98.7 | ⋯<br>⋯ | ⋯ | 1.60<br>1.55 |
| Poisson point distribution system | 1.0<br>0.1 | 10 000<br>94.1 | ⋯<br>⋯ | ⋯ | 1.98<br>1.93 |
| Lennard–Jones fluid system | 1.0<br>0.1 | 10 000<br>100.4 | ⋯<br>⋯ | ⋯ | 1.51<br>1.56 |
| Yukawa fluid system | 1.0<br>0.1 | 10 000<br>101.1 | ⋯<br>⋯ | ⋯ | 1.59<br>1.61 |

### B. Cropping ratio ($c$) dependence

We examine how spectral and real-space methods for detecting hyperuniformity depend on the crop ratio using the hyperuniform Gaussian pair statistics[56] as a representative system. Configurations with $N = 10,000$ particles were cropped with $c \in \{0.1, 0.2, 0.3, 0.4, 0.5, 0.6, 0.7, 0.8, 0.9, 1.0\}$, and each cropped configuration was rescaled to the target number density, as described in Sec. II C. For each $c$, we compute the structure factor $S(k)$, local number variance $\sigma_N^2(R)$, and pair correlation function $g_2(r)$, averaged over 20 configurations. We then extract exponents $\alpha$ and $\beta$ by fitting.

Figure 3 presents $S(k)$, $\sigma_N^2(R)$, and $g_2(r)$ after cropping, while Table III reports the fitted exponents for each crop ratio $c$. The uncropped configurations ($c = 1.0$) recover the theoretical small-$k$ scaling ($\alpha \approx 2.04$), whereas all cropped datasets ($c \leq 0.9$) exhibit substantially reduced $\alpha$ values in the small range of $\alpha \approx 1.56 \sim 1.65$. By contrast, the real-space exponent $\beta$ remains close to its theoretical value of 1 for the entire $c$-range (Table III), indicating that the class I surface-area scaling of $\sigma_N^2(R)$ is preserved even when the fitted spectral slope is diminished. For all $c$, the observations of $\alpha > 1$ and $\beta \approx 1$ together suggest that hyperuniformity with class I is preserved.

It is interesting to observe that moving from the full system to any $c < 1$ lowers the fitted $\alpha$, but there is no clear monotonic relationship between $\alpha$ and $c$, as the cropped $\alpha$ values remain concentrated in a relatively narrow range. Such a systematic reduction in $\alpha$ results from a systematic elevation in $S(k)$ in the smallest $k$ bins, which is also at roughly the same magnitude across different $c$ values.

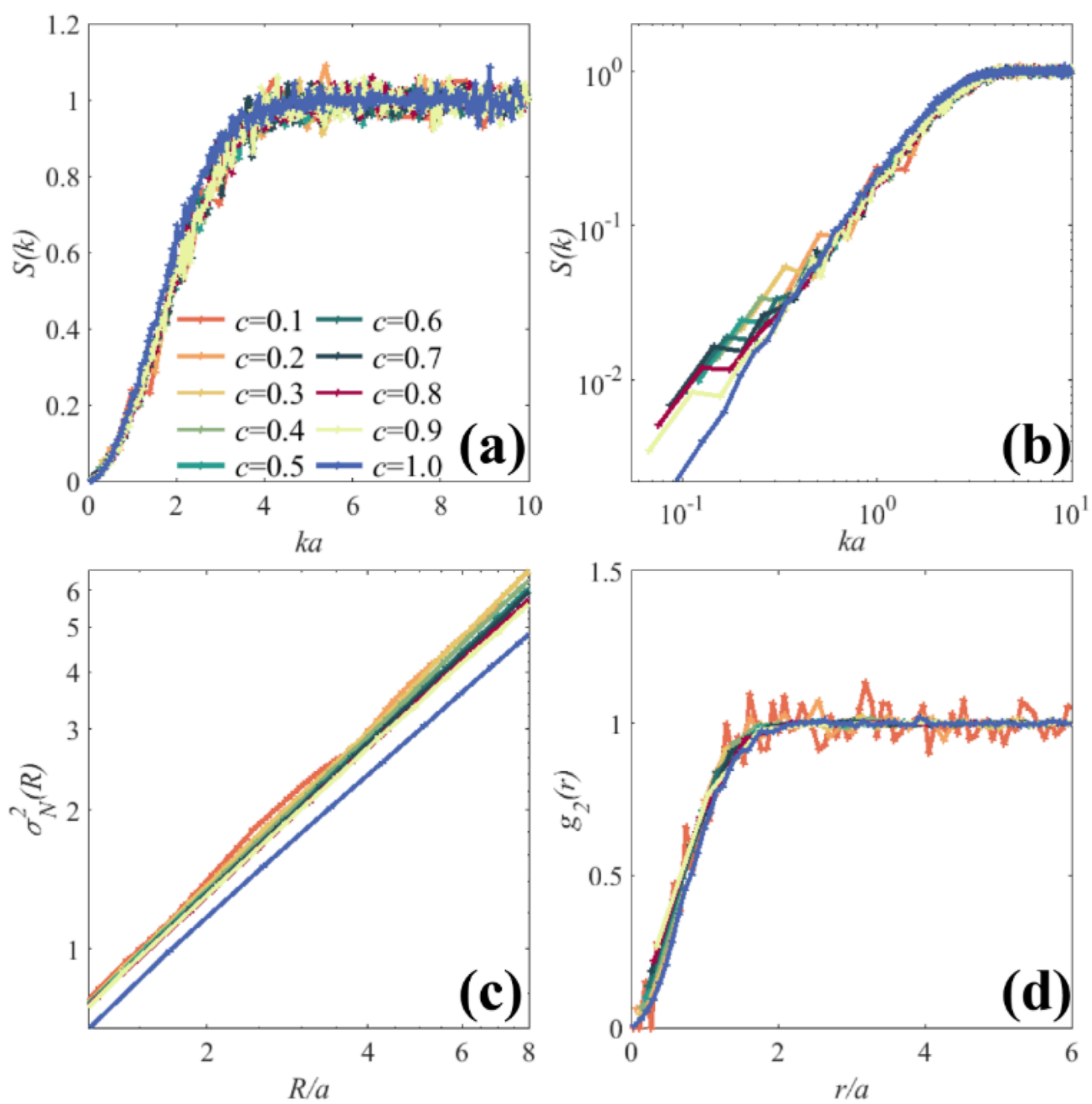

**FIG. 3.** Spectral and real-space quantities for detecting hyperuniformity for the Gaussian pair statistic system with various cropping ratios. Here, $a$ denotes the ensemble-average nearest-neighbor distance. Panel (a) shows the structure factor $S(k)$ on a linear scale. Panel (b) shows the same $S(k)$ data on a log–log scale. Panel (c) presents the local number variance $\sigma_N^2(R)$ on a log–log scale. Panel (d) displays the pair correlation function $g_2(r)$.

The primary mechanism behind this effect is the loss of true long-wavelength modes after spatial truncation, compounded by changes in the discrete-$k$ grid and by radial-rebinning artifacts. In particular, spatial truncation removes the longest wavelength modes, which increases the relative contribution of intermediate-wavelength fluctuations to the lowest measured bins. The rescaling step changes the discrete $k$-grid so that radial rebinning can place the smallest $\boldsymbol{k}$ vectors in configuration-dependent bins. This observation motivates us to study the radial binning effect in Subsection III C.

In summary, within a reasonable range of crop ratios, the overall structural hyperuniformity is retained, but spatial truncation induces a systematically elevated $S(k)$ in the smallest $k$ bins. The magnitude of this effect shows little dependence on $c$. Spatial truncation tends to reduce the apparent degree of disordered hyperuniformity as measured by the fitted $\alpha$, but this reduction is not sensitive to the exact choice of $c$ and does not change the asymptotic hyperuniform class.

**TABLE III.** Results for the hyperuniform Gaussian pair statistics systems at various crop ratios. For each crop ratio $c$, the average number of particles $\bar{N}_{crop}$, the fitted hyperuniformity exponent $\alpha$, and the fitted real-space exponent $\beta$ are listed.

| $c$ | $\bar{N}_{crop}$ | $\alpha$ | $\beta$ |
|---|---|---|---|
| 0.1 | 100.5 | 1.59 | 1.00 |
| 0.2 | 400.1 | 1.56 | 0.97 |
| 0.3 | 897.7 | 1.58 | 1.01 |
| 0.4 | 1601.1 | 1.60 | 1.00 |
| 0.5 | 2501.0 | 1.61 | 1.03 |
| 0.6 | 3595.2 | 1.62 | 1.04 |
| 0.7 | 4896.8 | 1.57 | 1.02 |
| 0.8 | 6394.2 | 1.62 | 1.03 |
| 0.9 | 8098.4 | 1.65 | 1.02 |
| 1.0 | 10 000 | 2.04 | 0.99 |

### C. Radial binning ($m$) and mitigation of small-$k$ fluctuations

To address the small-$k$ error induced by spatial truncation, we study the effect of radial smoothing (binning) in the reciprocal space, using the hyperuniform Gaussian pair statistics ensemble[56] at $c = 0.5$ as a representative case. The radial bin width is $k_{Bin} = m \cdot 2\pi/L$ with $L$ denoting the side length of the periodic box. Notably, wavevector quantization under periodic boundary conditions makes $m \geq 1.0$ necessary.[66,67] We examine $m \in \{1.0,\ 1.2,\ 1.5,\ 2.0,\ 2.5,\ 3.0,\ 3.5,\ 4.0\}$. For each $m$, we calculated the configuration-averaged structure factor $S(k)$ and extract the small-$k$ hyperuniformity exponent $\alpha$. The real-space exponent $\beta$ is obtained from the asymptotic scaling of the configuration-averaged number variance $\sigma_N^2(R)$ and, therefore, does not vary with $m$. Figure 4 displays the $S(k)$ curves for the tested $m$ values, while Table IV reports the observed minimum plotted wave number $k_{\min}$ and fitted $\alpha$.

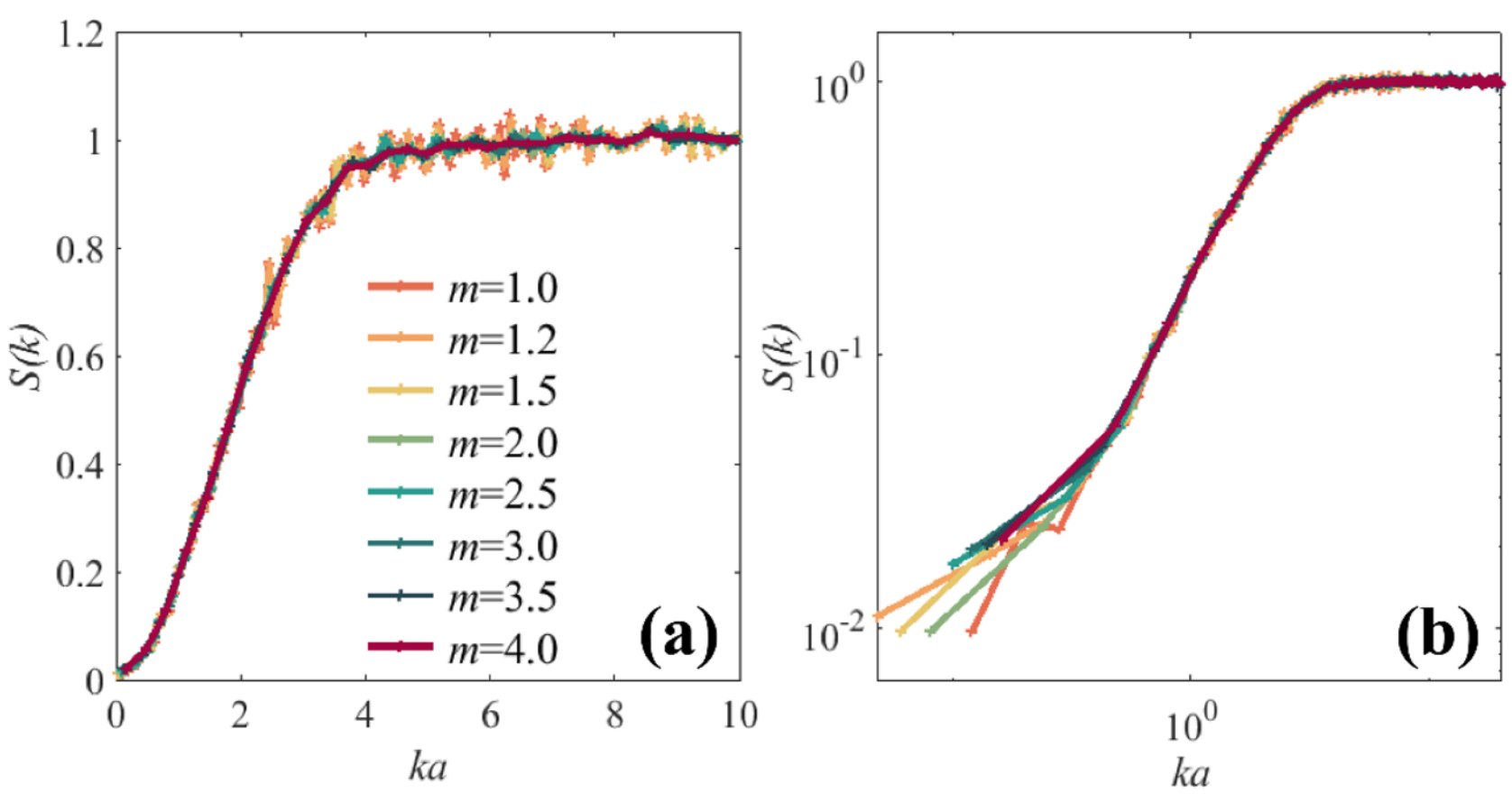

**FIG. 4.** Structure factor $S(k)$ for the hyperuniform Gaussian pair-statistics ensemble (with crop ratio $c = 0.5$) plotted for different radial binning parameters $m$. Here, $a$ denotes the ensemble-average nearest-neighbor distance. Panel (a) shows $S(k)$ on a linear scale, while panel (b) shows the same data on a log–log scale to present the small-$k$ behavior.

From the numerical results, we draw several practical observations. First, moderate $m$ produces visibly smoother $S(k)$ and reduces the prominence of fluctuations, while leaving the fitted $\alpha$ unchanged. Second, the minimum plotted wave number $k_{\min}$ does not vary monotonically with $m$. This effect results from discrete radial binning. When $m = 1.0$, the first radial bin may contain no $k$ vectors for a given configuration, while $k_{\min}$ appears in the second bin. Slightly increasing $m$ can place at least one $k$ vector into the first bin and thereby produce a smaller observed $k_{\min}$. This behavior shows that the selected $m$ can influence which small $k$ points are available for plotting and fitting. Third, very large values of $m$ increase $k_{\min}$ substantially and reduce the number of data points in the small $k$ region. This data loss can obscure information that is critical for determining the small $k$ slope and may lower the fitted $\alpha$.

Taken together, the optimal choice of $m$ must be determined from a trade-off between visual smoothness and retention of small-$k$ information. However, it can also depend on the system, specific configurations, and the severity of fluctuations. Therefore, we recommend moderate radial coarse-graining and suggest $1.0 < m \leq 3.0$ as a practical starting range. One should determine the smallest $m$ that effectively suppresses fluctuations while keeping $k_{\min}$ as small as possible by performing an $m$-scan. As we demonstrated in Sec. III B, the binning guideline we suggested here is applicable for two-dimensional systems with a wide range of $N_{crop} \in [10^2,\ 10^4]$. For datasets outside the range considered here, a wider $m$-scan may be necessary. Finally, when recovering theoretical spectral exponents from finite, truncated data, one should validate spectral fits using the local number variance scaling.

**TABLE IV.** Dependence of fitted exponents on the radial-binning parameter $m$ for the hyperuniform Gaussian pair-statistics ensemble at $c = 0.5$. The columns present the radial binning parameter $m$, the observed minimum plotted wave number (where $a$ is the ensemble-average nearest-neighbor distance), the fitted small-$k$ hyperuniformity exponent $\alpha$ obtained from the configuration-averaged $S(k)$, and the real-space exponent $\beta$ obtained from $\sigma_N^2(R)$.

| $m$ | $ka_{\min}$ | $\alpha$ | $\beta$ |
|---|---|---|---|
| 1.0 | 0.120 775 | 1.61 | |
| 1.2 | 0.048 310 | 1.58 | |
| 1.5 | 0.060 387 | 1.56 | |
| 2.0 | 0.080 516 | 1.63 | |
| 2.5 | 0.100 646 | 1.52 | 1.03 |
| 3.0 | 0.120 775 | 1.50 | |
| 3.5 | 0.140 904 | 1.45 | |
| 4.0 | 0.161 033 | 1.43 | |

## D. Application to experimental biological systems

To test the effects of spatial truncation and reciprocal-space radial binning in real-life biological systems, we applied the same analysis protocol to two experimental biological systems previously known to exhibit disordered hyperuniformity: avian photoreceptor patterns[39] and looped leaf vein networks.[43] The photoreceptor dataset contains five receptor types, each exhibiting hyperuniformity. Thus, we selected random square spatial truncations with type-specific crop ratios so that the resulting cropped particle counts $N_{crop}$ are comparable across receptor types (red, green, blue, violet, double, and the combined overall population). The cropped configurations were rescaled to a fixed number density as described in Sec. II C. For the leaf vein networks, since the configurations were originally obtained by cropping from whole leaves and already exhibited pronounced finite-size effects, we analyzed the available full configurations without further cropping.

Figures 5(a), 5(b), 6(a), and 6(b) show the configuration-averaged structure factors $S(k)$ on linear and log scales for the cropped avian photoreceptor patterns and for the looped leaf vein networks, respectively. As in the controlled simulation systems, the cropped configurations exhibit varying degrees of small-$k$ fluctuations. Guided by the findings in Sec. III C, we recomputed $S(k)$ using a reciprocal-space radial binning parameter $m = 2.0$. The corresponding spectra are shown in Figs. 5(c), 5(d), 6(c), and 6(d). The $m = 2.0$ results lose some fine detail, but are visually smoother in the small-$k$ region as the fluctuations are largely removed. Table V summarizes the experimental biological results. For each sample, we list the system and species identifier, the crop ratio $c$ applied to the original dataset, the average cropped particle count $\overline{N}_{crop}$, and the fit-

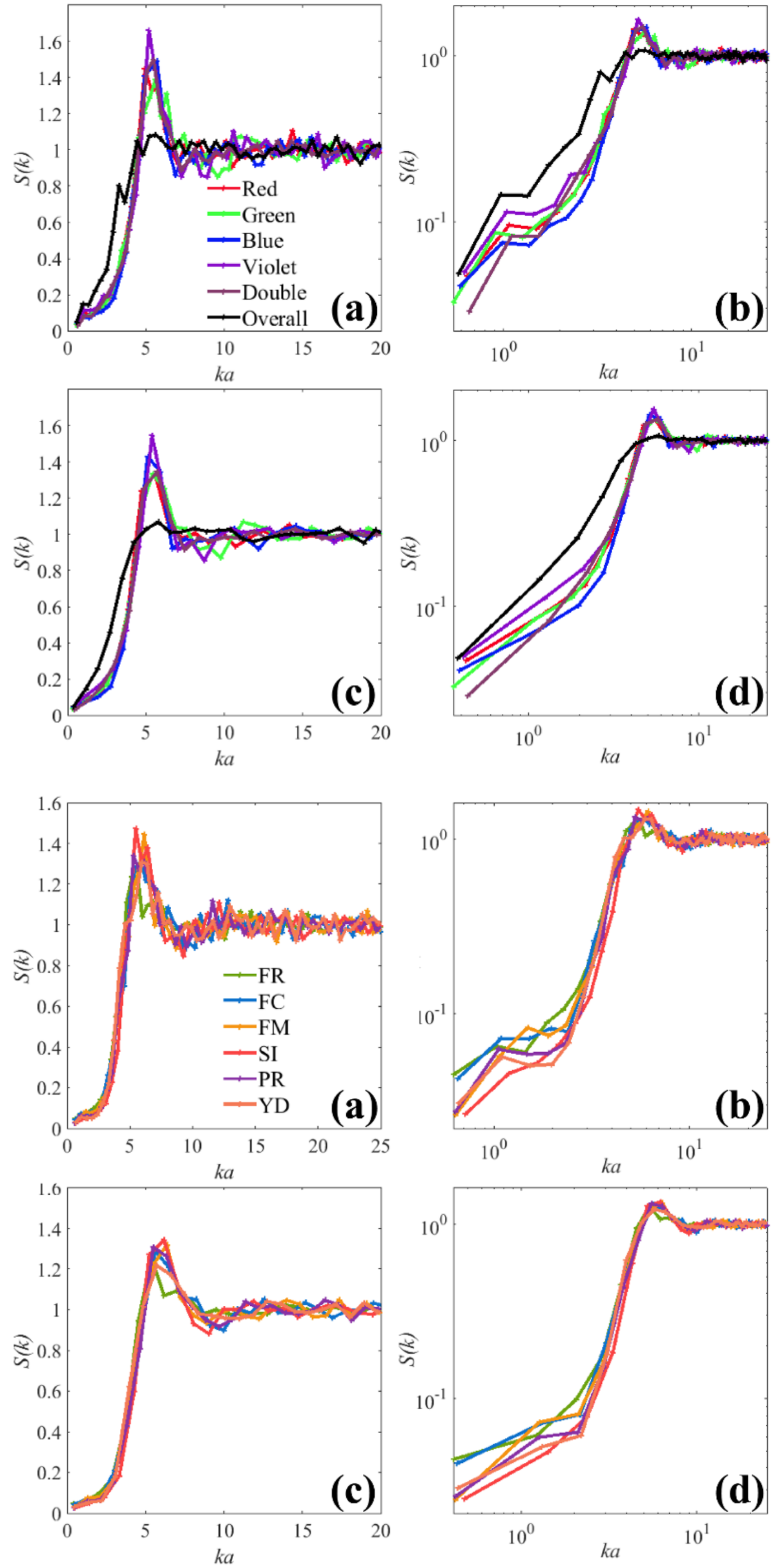

**FIG. 5.** Structure factors for cropped avian photoreceptor patterns. Panels [(a) and (b)] show $S(k)$ on a linear and a log–log scale computed with the fundamental radial bin width $m = 1.0$. Panels [(c) and (d)] show the same data recomputed with $m = 2.0$ on a linear and a log–log scale.

**FIG. 6.** Structure factors for looped leaf vein networks. Panels [(a) and (b)] show $S(k)$ on a linear and a log–log scale using $m = 1.0$. Panels [(c) and (d)] show the same spectra computed with $m = 2.0$ on a linear and a log–log scale.

**TABLE V.** Fitted hyperuniformity exponents $\alpha$ for experimental biological systems, obtained from structure factors $S(k)$ using two radial binning parameter choices $m = 1.0$ and $m = 2.0$. The columns list the system and species identifier, the crop ratio $c$ applied to the original field of view, the average cropped particle count $\bar{N}_{crop}$, and the fitted $\alpha$ values for $m = 1.0$ and $m = 2.0$. For the avian photoreceptor rows, the reported crop ratios are class-specific and chosen so that $N_{crop}$ is comparable across receptor classes; leaf vein rows report results on the original fields of view (no additional cropping, $c = 1.0$).

| System | Species | $c$ | $\bar{N}_{crop}$ | $\alpha$ ($m = 1.0$) | $\alpha$ ($m = 2.0$) |
|---|---|---|---|---|---|
| Avian photoreceptor patterns | Red | 0.5 | 134.0 | 1.35 | 1.30 |
| | Green | 0.5 | 189.2 | 1.21 | 1.21 |
| | Blue | 0.6 | 166.1 | 1.21 | 1.22 |
| | Violet | 0.7 | 143.9 | 1.20 | 1.19 |
| | Double | 0.3 | 128.0 | 1.37 | 1.40 |
| | Overall | 0.2 | 135.1 | 1.29 | 1.25 |
| Leaf vein networks | Ficus religiosa (FR) | 1.0 | 144.6 | 0.64 | 0.70 |
| | Ficus caulocarpa (FC) | 1.0 | 135.2 | 0.65 | 0.69 |
| | Ficus microcarpa (FM) | 1.0 | 142.2 | 0.76 | 0.79 |
| | Smilax indica (SI) | 1.0 | 118.4 | 0.78 | 0.77 |
| | Populus rotundifolia (PR) | 1.0 | 146.2 | 0.69 | 0.67 |
| | Yulania denudata (YD) | 1.0 | 130.0 | 0.63 | 0.61 |

ted hyperuniformity exponent $\alpha$ obtained from $S(k)$ for $m = 1.0$ and $m = 2.0$.

The results of experimental biological systems support two practical conclusions. First, moderate $m$ removes spurious small-$k$ fluctuations and improves the visual clarity of $S(k)$ while sacrificing only limited spectral detail. Second, this smoothing does not materially change the fitted hyperuniform exponent $\alpha$. The values of $\alpha$ obtained with $m = 1.0$ and $m = 2.0$ are similar and lead to the same hyperuniform classification. Therefore, experimental biological tests corroborate the conclusions we developed in simulations: Moderate radial pooling ($m = 2.0$) provides a low-cost and effective remedy for fluctuations in the smallest-$k$ bins caused by spatial truncation and finite sampling, and it does so without substantially changing the small-$k$ decline that underlies disordered hyperuniform classification.

## IV. CONCLUSIONS AND DISCUSSION

We examined how finite-window sampling and reciprocal-space radial binning affect the detection of hyperuniformity in disordered systems. Our study combined thirteen representative two-dimensional controlled simulation systems with two experimental biological systems. We focused on the experimentally relevant regime of the limited number of configurations ($N_c = 20$). Therefore, the conclusions, as we summarize below, have direct implications for empirical studies of hyperuniformity.

First, random spatial truncation does not change the qualitative hyperuniform classification of a system. Spatial truncation does reduce the degree of $S(k)$ suppression at the longest accessible wavelengths. This causes a significant downward shift in the fitted exponent $\alpha$ for large $\alpha$, but the shift diminishes and becomes negligible for $\alpha \leq 1$. Therefore, spatial truncation does not change the hyperuniformity class. The corresponding real-space exponent $\beta$ for the local number variance remains comparatively stable. Spatial truncation also preserves short-range structural signatures and the overall shapes of $S(k)$ and $g_2(r)$ curves. In the supplementary material, we demonstrate that the *systematic* shift in $\alpha$ is caused by spatial truncation, while the limited number of configurations ($N_c = 20$) causes *random* errors in $\alpha$. In the Appendix, we develop an analytical theory that explains (1) why is the shift negligible for $\alpha \leq 1$ and (2) why do we observe $\alpha \approx 2$ for cropped stealthy systems.

Second, spatial truncation commonly induces configuration-dependent systematic errors and fluctuations in the smallest measured $k$ bins. Two related factors drive this outcome. First, spatial truncation removes the longest-wavelength modes. This removal raises the relative contribution of intermediate-wavelength fluctuations. Second, rescaling after cropping also assigns a different effective box size $L_{crop}$ to each configuration. Different configurations can, therefore, contribute to different bins in calculating $S(k)$. These combined effects tend to accentuate small-$k$ fluctuations in the cropped data. As a result, small-$k$ fits become more sensitive to configuration noise and to the choice of the binning protocol.

Third, moderate reciprocal-space radial binning is an effective and low-cost remedy for sampling-induced fluctuations in the smallest-$k$ bins. Increasing the radial bin width, parameterized by $m$, smooths out configuration-dependent fluctuations while preserving the overall small-$k$ trend that determines hyperuniform classification. Fine-tuning $m$ is required because overly large $m$ removes smallest $k$ data points and reduces the overall number of data points available for fitting, which may bias the fitted $\alpha$. Based on our tests on simulational and experimental datasets, we recommend moderate radial coarse-graining with $1.0 < m \leq 3.0$, while our datasets produce optimal results near $m = 2.0$. Within this range, the fitted $\alpha$ is roughly constant, and the hyperuniform class inferred is not changed. Our quantitative binning range of $m$ was calibrated using two-dimensional systems with $N_{crop} \in [10^2,\ 10^4]$; finding whether our choice for $m$ applies to three dimensions (and a broader range of $N_{crop}$) is an important direction for future work.

In practice, we advise the following: Perform an $m$-scan and choose the smallest $m$ that visibly suppresses spurious fluctuations while keeping $k_{\min}$ as small as possible. Cross-check spectral inferences with the local number variance scaling. Where feasible, increase effective sampling by averaging over more configurations.

Future work that would further improve practical methodology for detecting hyperuniformity includes a systematic study of non-square and tapered windows, extension to three dimensions, and a generalization of our study to anisotropic or strongly inhomogeneous systems. These directions will help refine best practices for robust detection and quantification of hyperuniformity in diverse experimental contexts.

## SUPPLEMENTARY MATERIAL

The supplementary material contains results for different numbers of configurations ($N_c$). They demonstrate that for the Gaussian pair statistics system (class I), spatial truncation causes a *systematic* downward shift in the fitted $\alpha$, while the limited number of configurations ($N_c = 20$ in the main text) causes a *random* error.

## ACKNOWLEDGMENTS

We thank Carlo Vanoni for helpful discussions. This work was supported by the National Natural Science Foundation of China (Grant Nos. 11274200 and 12274255), the Natural Science Foundation of Shandong Province, China (Grant No. ZR2022MA055), the foundation with Grant No. 2022NS189 from XYIAS, and City University of Hong Kong (Grant No. 9610532). The authors sincerely appreciate the anonymous reviewers for devoting their precious time to reviewing our work.

## AUTHOR DECLARATIONS

### Conflict of Interest

The authors have no conflicts to disclose.

### Author Contributions

Y.L. and X.L. contributed equally to this work.

**Yuan Liu (刘源)**: Data curation (equal); Software (equal). **Xurui Li (李旭瑞)**: Data curation (equal); Software (equal). **Jianxiang Tian (田建祥)**: Data curation (equal); Investigation (equal); Project administration (equal); Writing – original draft (equal); Writing – review & editing (equal). **Xunwang Yan (闫循旺)**: Data curation (equal); Investigation (equal); Writing – review & editing (equal). **Ge Zhang (张格)**: Data curation (equal); Investigation (equal); Writing – review & editing (equal).

## DATA AVAILABILITY

The configurations studied in this research can be downloaded at https://www.dropbox.com/scl/fo/yyai5qq6cfjaafr7f3p19/AHELFX6ZyORuJB44UFdtulE?rlkey=a6w3wzsvrac11nkv6lykh7iia&st=xvuwvcoc&dl=0.

## APPENDIX: ANALYTICAL EXPLANATION TO THE NUMERICAL RESULTS FOR FINITE OBSERVATION WINDOWS

Here, we analytically explain our numerical observations, as follows:

1. Cropping causes very little change in $\alpha$ for class III ($0 < \alpha < 1$) systems.
2. Cropped stealthy systems exhibit $\alpha = 2$.

To interpret our numerical results, it is useful to develop an analytical theory on cropping. For a point configuration with particle positions $\{\boldsymbol{r}_j\}_{j=1}^{N}$, the microscopic number density is

$$\rho(\boldsymbol{r}) = \sum_{j=1}^{N} \delta(\boldsymbol{r} - \boldsymbol{r}_j). \tag{A1}$$

Extracting a finite field of view is then exactly equivalent to the operation of multiplying the parent density by a window function $W_{r_0}(\boldsymbol{r})$ that equals to one inside the observation region and zero outside, so the windowed (i.e., cropped) density can be written as[69,70]

$$\rho_{sub}(\boldsymbol{r}) = \rho(\boldsymbol{r}) W_{r_0}(\boldsymbol{r}), \tag{A2}$$

where $r_0$ denotes the window center.[71,72] A translated window differs from a window centered at the origin only by a phase factor in Fourier space,

$$\tilde{W}_{r_0}(\boldsymbol{k}) = e^{-i\boldsymbol{k}\cdot\boldsymbol{r}_0} \tilde{W}(\boldsymbol{k}). \tag{A3}$$

Using the Fourier transform convention,

$$\tilde{f}(\boldsymbol{k}) = \int d\boldsymbol{r} f(\boldsymbol{r}) e^{-i\boldsymbol{k}\cdot\boldsymbol{r}}, \quad f(\boldsymbol{r}) = \frac{1}{(2\pi)^2} \int d\boldsymbol{k} \tilde{f}(\boldsymbol{k}) e^{i\boldsymbol{k}\cdot\boldsymbol{r}}, \tag{A4}$$

multiplication in real space becomes convolution in reciprocal space.[69,70] Inserting the inverse Fourier representation of $\rho(\boldsymbol{r})$ into $\tilde{\rho}_{sub}(\boldsymbol{k}) = \int d\boldsymbol{r} \rho(\boldsymbol{r}) W_{r_0}(\boldsymbol{r}) e^{-i\boldsymbol{k}\cdot\boldsymbol{r}}$ and exchanging integration order yields the exact convolution form,

$$\tilde{\rho}_{sub}(\boldsymbol{k}) = \frac{1}{(2\pi)^2} \int d\boldsymbol{q} \tilde{\rho}(\boldsymbol{q}) \tilde{W}_{r_0}(\boldsymbol{k} - \boldsymbol{q}). \tag{A5}$$

This already shows that the Fourier modes measured from a truncated configuration are not a clean sampling of the parent long-wavelength modes; instead, they involve a local mixing of spectral content over a bandwidth controlled by $\tilde{W}$.[69,70,72] To connect this directly to the structure factor, consider the quadratic spectral quantity $\langle |\tilde{\rho}_{sub}(\boldsymbol{k})|^2 \rangle$. Substituting Eq. (A5) and expanding the modulus gives a double integral with the window-center phase factors made explicit via Eq. (A3),

$$\begin{aligned} \langle |\tilde{\rho}_{sub}(\boldsymbol{k})|^2 \rangle = \frac{1}{(2\pi)^4} \int d\boldsymbol{q} \int d\boldsymbol{q}'^{\langle \tilde{\rho}(\boldsymbol{q}) \tilde{\rho}^*(\boldsymbol{q}') \rangle} \tilde{W} \\ (\boldsymbol{k} - \boldsymbol{q}) \tilde{W}^*(\boldsymbol{k} - \boldsymbol{q}') \left\langle e^{-i(\boldsymbol{q} - \boldsymbol{q}')\cdot\boldsymbol{r}_0} \right\rangle_{\boldsymbol{r}_0}. \end{aligned} \tag{A6}$$

If the observation window center $\boldsymbol{r}_0$ is uniformly sampled (random cropping), the average over $\boldsymbol{r}_0$ suppresses cross terms, since

$$\left\langle e^{-i(\boldsymbol{q} - \boldsymbol{q}'\cdot\boldsymbol{r}_0)} \right\rangle_{\boldsymbol{r}_0} \propto (2\pi)^2 \delta(\boldsymbol{q} - \boldsymbol{q}'), \tag{A7}$$

and so, Eq. (A6) collapses into a single-$\boldsymbol{q}$ integral,

$$\left\langle |\tilde{\rho}_{sub}(\boldsymbol{k})|^2 \right\rangle \propto \int d\boldsymbol{q} \left\langle |\tilde{\rho}(\boldsymbol{q})|^2 \right\rangle \left| \tilde{W}(\boldsymbol{k}-\boldsymbol{q}) \right|^2 . \tag{A8}$$

Introducing structure-factor-like normalizations for the parent and the subsystem,

$$S(\boldsymbol{k}) = \frac{1}{N} \left\langle |\tilde{\rho}(\boldsymbol{k})|^2 \right\rangle, \ \ S_{sub}(\boldsymbol{k}) = \frac{1}{\langle N_{sub} \rangle} \left\langle |\tilde{\rho}_{sub}(\boldsymbol{k})|^2 \right\rangle, \tag{A9}$$

and using $\langle N_{sub} \rangle = \rho V_{sub}$ for a homogeneous system, one obtains the relation that describes the physical effect of cropping,

$$S_{sub}(\boldsymbol{k}) \propto \int d\boldsymbol{q} S(\boldsymbol{q}) K(\boldsymbol{k}-\boldsymbol{q}), \tag{A10}$$

where the proportionality is due to convention-dependent normalization factors (Fourier conventions and window volume factors) and $K(\boldsymbol{k}-\boldsymbol{q}) = \left| \tilde{W}(\boldsymbol{k}-\boldsymbol{q}) \right|^2$. This equation shows that the subsampled structure factor is the convolution of the original structure factor and a kernel function $K$, which encodes spectral leakage from $\boldsymbol{q}$ into the measured $\boldsymbol{k}$. This windowed-spectrum picture is standard in discrete Fourier spectral estimation and in the spectral analysis of spatial point processes.[69–72] For a square window of side $L_{sub}$ centered at the origin, the Fourier transform factorizes into sinc functions,

$$\tilde{W}(\boldsymbol{k}) = L_{sub}^2 \, sinc\left(\frac{k_x L_{sub}}{2}\right) sinc\left(\frac{k_y L_{sub}}{2}\right), sinc(x) = \frac{\sin x}{x}. \tag{A11}$$

The kernel has long algebraic tails and strong anisotropy.[69,70] In particular, if both $k_x$ and $k_y$ are large, then $K(\boldsymbol{k}) = \left| \tilde{W}(\boldsymbol{k}) \right|^2 \propto k_x^{-2} k_y^{-2} \propto |\boldsymbol{k}|^{-4}$. However, if $k_x$ is small, then $sinc\left(\frac{k_x L_{sub}}{2}\right) \approx 1$, and $K(\boldsymbol{k}) \propto k_y^{-2} \propto |\boldsymbol{k}|^{-2}$. Similarly, if $k_y$ is small, the kernel also decays quadratically.

It is noteworthy that we are typically in the large-$k_x$ and large-$k_y$ regimes, unless $k_x$ or $k_y$ is zero. This is because the wavenumber resolution for the truncated system is

$$k_{\min} \sim \frac{2\pi}{L_{sub}}. \tag{A12}$$

Therefore, $k_x$ and $k_y$ are comparable to $1/L_{sub}$ for the first few bins and are much larger than $1/L_{sub}$ for later bins. At a length scale much smaller than $L_{sub}$ ($\boldsymbol{k}$-scale much larger than $1/L_{sub}$), the kernel is a fast-oscillating and decaying function rather than a constant.

If the system before truncation has a power-law structure factor, $S(\boldsymbol{q}) = |\boldsymbol{q}|^{\alpha}$, does the convolution converge? Substituting the power law into Eq. (A10) gives $S_{sub}(\boldsymbol{0}) \propto \int d\boldsymbol{q} |\boldsymbol{q}|^{\alpha} K(-\boldsymbol{q})$. As we discussed, the kernel $K(-\boldsymbol{q})$ decays as $|\boldsymbol{q}|^{-4}$ overall, but in horizontal and vertical directions, it only decays as $|\boldsymbol{q}|^{-2}$. Therefore, for class III systems, $0 < \alpha < 1$, and the integrand decays faster than $|\boldsymbol{q}|^{-1}$ horizontally and vertically, leading to convergence. The fact that for class III systems, $S_{sub}$ converges even if the power law holds for infinitely large $\boldsymbol{q}$ has an important implication. In particular, it shows that in this convolution, the dominant contribution to $S_{sub}(\boldsymbol{k})$ is local (i.e., comes from $\boldsymbol{q} \approx \boldsymbol{k}$). This is because if the dominant contribution came from distant $\boldsymbol{q}$ vectors, then the integral would diverge since distant $\boldsymbol{q}$ vectors occupy an infinite volume. The locality of the convolution integral explains why truncation hardly changes $\alpha$ for $\alpha < 1$.

For class I and class II systems, $\alpha \geq 1$, and $S_{sub}(\boldsymbol{0}) \propto \int d\boldsymbol{q} |\boldsymbol{q}|^{\alpha} K(-\boldsymbol{q})$ diverges. In reality, we do not see $S_{sub}(\boldsymbol{0}) \to \infty$ since the power law, $S(\boldsymbol{q}) = |\boldsymbol{q}|^{\alpha}$, is not valid for large $\boldsymbol{q}$. Nevertheless, the fact that the difference between $S(\boldsymbol{q})$ and $|\boldsymbol{q}|^{\alpha}$ for large $\boldsymbol{q}$ matters implies that distant contributions are important in the convolution, unlike class III systems. Therefore, the exponent of $S_{sub}(\boldsymbol{q})$ and $S(\boldsymbol{q})$ can differ.

It is also possible to theoretically analyze the effect of truncation in stealthy hyperuniform systems. In the thermodynamic limit, a stealthy pattern exhibits a spectral gap,

$$S(\boldsymbol{k}) = 0 \ \text{for any} \ |\boldsymbol{k}| < K, \tag{A13}$$

so there is no intrinsic power-law scaling inside the gap. After truncation, Eq. (A10) implies that the measured spectrum in the nominal gap region is entirely leakage-driven, because only $|\boldsymbol{q}| \gtrsim K$ contributes,

$$S_{sub}(\boldsymbol{k}) \propto \int_{|\boldsymbol{q}| \gtrsim K} d\boldsymbol{q} S(\boldsymbol{q}) K(\boldsymbol{k}-\boldsymbol{q}), \ \ |\boldsymbol{k}| < K . \tag{A14}$$

To extract the small-$k$ behavior deep inside the gap, one expands the kernel for $|\boldsymbol{k}| \ll |\boldsymbol{q}|$ about $|\boldsymbol{k}| = 0$,

$$K(\boldsymbol{k}-\boldsymbol{q}) = K(\boldsymbol{q}) - k_a \partial_a K(\boldsymbol{q}) + \frac{1}{2} k_a k_b \partial_a \partial_b K(\boldsymbol{q}) + \mathcal{O}\left(k^3\right), \tag{A15}$$

where repeated indices are summed. Under the isotropy of $S(\boldsymbol{q})$ and symmetry of the windowing protocol (and, in practice, after angular averaging or radial binning), the term linear in $\boldsymbol{k}$ does not contribute because it is odd under $\boldsymbol{q} \to -\boldsymbol{q}$, so the leading scale-dependent correction is quadratic,

$$\int d\boldsymbol{q} S(\boldsymbol{q}) \partial_a K(\boldsymbol{q}) = 0 \ \ \text{(after angular averaging / symmetry)}. \tag{A16}$$

Since the linear term in Eq. (A15) does not contribute, a generic small-$k$ expansion for the windowed spectrum has the following form:

$$S_{sub}(k) = S_0 + Ck^2 + \mathcal{O}\left(k^4\right). \tag{A17}$$

Here, $S_0$ is a small but nonzero leakage floor introduced by windowing, and the leading $k$-dependent correction is quadratic. As we have explained, typical $k$ values in realistic settings vary between $k_{\min} = \frac{2\pi}{L_{sub}}$ and several times larger than $k_{\min}$. Therefore, the wavenumber scale is typically larger than the oscillating scale of the kernel, and $\frac{1}{2} k_a k_b \partial_a \partial_b K(\boldsymbol{q})$ is much larger than $K(\boldsymbol{q})$. Thus, the quadratic term dominates, which explains the observed $S_{sub}(k) \sim k^2$ for truncated stealthy systems.

We have numerically verified that the quadratic term in Eq. (A17), indeed, dominates. To do so, we have fitted $S_{sub}(k) = S_0 + Ck^2$ to the numerically computed structure factor of cropped ($c = 0.1$) stealthy systems with $\chi = 0.2$ and $\chi = 0.49$. Our results are presented in Tables VI and VII. Indeed, the quadratic term is 1–2 orders of magnitude larger than the constant term. The two cropped stealthy systems have slightly different $k_{\min}$; this is due to fluctuations in the rescaling step (details in Sec. II C).

**TABLE VI.** Fitted constant term and second-order term for cropped stealthy systems with $\chi = 0.2$.

| $k$ | $S_0$ | $Ck^2$ | $Ck^2/S_0$ |
|---|---|---|---|
| $k_{\min} = 1.149\,848$ | 0.001 025 | 0.018 938 | 18.48 |
| 1.916 415 | 0.001 025 | 0.052 604 | 51.32 |
| 2.682 980 | 0.001 025 | 0.103 146 | 100.63 |
| 3.449 547 | 0.001 025 | 0.170 509 | 166.35 |
| 4.216 110 | 0.001 025 | 0.254 713 | 248.50 |

**TABLE VII.** Same as Table VI, except for $\chi = 0.49$.

| $k$ | $S_0$ | $Ck^2$ | $Ck^2/S_0$ |
|---|---|---|---|
| $k_{\min} = 1.156\,653$ | 0.000 364 | 0.006 741 | 18.52 |
| 1.927 759 | 0.000 364 | 0.018 724 | 51.44 |
| 2.698 860 | 0.000 364 | 0.036 699 | 100.82 |
| 3.469 965 | 0.000 364 | 0.060 666 | 166.67 |
| 4.241 069 | 0.000 364 | 0.090 624 | 248.97 |